\documentstyle[12pt,epsfig]{article}
\advance\textheight by \topskip
\begin{document}
%\baselineskip 1.1truecm
\tolerance=100000
\thispagestyle{empty}
\setcounter{page}{0}

\newcommand{\be}{\begin{equation}}
\newcommand{\ee}{\end{equation}}
\newcommand{\br}{\begin{eqnarray}}
\newcommand{\er}{\end{eqnarray}}
\newcommand{\ba}{\begin{array}}
\newcommand{\ea}{\end{array}}
\newcommand{\bi}{\begin{itemize}}
\newcommand{\ei}{\end{itemize}}
\newcommand{\bn}{\begin{enumerate}}
\newcommand{\en}{\end{enumerate}}
\newcommand{\bc}{\begin{center}}
\newcommand{\ec}{\end{center}}
\newcommand{\ul}{\underline}
\newcommand{\ol}{\overline}
\newcommand{\ar}{\rightarrow}
\newcommand{\sm}{${\cal {SM}}$}
\newcommand{\susy}{{{SUSY}}}
\newcommand{\Dir}{\kern -6.4pt\Big{/}}
\newcommand{\Dirin}{\kern -12.4pt\Big{/}\kern 4.4pt}
\newcommand{\DGir}{\kern -6.0pt\Big{/}}
\def\pp{\ifmmode{pp} \else{$pp$} \fi}
\def\CC{\ifmmode{{\it C.C.}} \else{$\mbox{\it C.C.}$} \fi}
\newcommand{\bqbqH}{\ifmmode{bU\ar bDH^+} 
                   \else{$bU\ar bDH^+$}\fi}
\newcommand{\bqbqW}{\ifmmode{bU\ar bDW^+} 
                   \else{$bU\ar bDW^+$}\fi}
\newcommand{\qqbbW}{\ifmmode{U\bar D\ar b\bar bW^+} 
                   \else{$U\bar D\ar b\bar bW^+$}\fi}
\newcommand{\bqbqtn}{\ifmmode{bU\ar bD\tau^+\nu_\tau} 
                   \else{$bU\ar bD\tau^+\nu_\tau$}\fi}
\newcommand{\qqbbtn}{\ifmmode{U\bar D\ar b\bar b\tau^+\nu_\tau} 
                   \else{$U\bar D\ar b\bar b\tau^+\nu_\tau$}\fi}
\newcommand{\Htn}{\ifmmode{H^\pm\ar \tau\nu_\tau}
                   \else{$H^\pm\ar \tau\nu_\tau$}\fi}
\newcommand{\Wtn}{\ifmmode{W^\pm\ar \tau\nu_\tau}
                   \else{$W^\pm\ar \tau\nu_\tau$}\fi}
\def\mssm{\ifmmode{{\cal {MSSM}}}\else{${\cal {MSSM}}$}\fi}
\def\MH{\ifmmode{{M_{H}}}\else{${M_{H}}$}\fi}
\def\Mh{\ifmmode{{M_{h}}}\else{${M_{h}}$}\fi}
\def\MA{\ifmmode{{M_{A}}}\else{${M_{A}}$}\fi}
\def\MHpm{\ifmmode{{M_{H^\pm}}}\else{${M_{H^\pm}}$}\fi}
\def\tb{\ifmmode{\tan\beta}\else{$\tan\beta$}\fi}
\def\ctb{\ifmmode{\cot\beta}\else{$\cot\beta$}\fi}
\def\ta{\ifmmode{\tan\alpha}\else{$\tan\alpha$}\fi}
\def\cta{\ifmmode{\cot\alpha}\else{$\cot\alpha$}\fi}
\def\tba{\ifmmode{\tan\beta=1.5}\else{$\tan\beta=1.5$}\fi}
\def\tbb{\ifmmode{\tan\beta=30}\else{$\tan\beta=30.$}\fi}
\def\cab{\ifmmode{c_{\alpha\beta}}\else{$c_{\alpha\beta}$}\fi}
\def\sab{\ifmmode{s_{\alpha\beta}}\else{$s_{\alpha\beta}$}\fi}
\def\cba{\ifmmode{c_{\beta\alpha}}\else{$c_{\beta\alpha}$}\fi}
\def\sba{\ifmmode{s_{\beta\alpha}}\else{$s_{\beta\alpha}$}\fi}
\def\ca{\ifmmode{c_{\alpha}}\else{$c_{\alpha}$}\fi}
\def\sa{\ifmmode{s_{\alpha}}\else{$s_{\alpha}$}\fi}
\def\cb{\ifmmode{c_{\beta}}\else{$c_{\beta}$}\fi}
\def\sb{\ifmmode{s_{\beta}}\else{$s_{\beta}$}\fi}

\def\Ord{\buildrel{\scriptscriptstyle <}\over{\scriptscriptstyle\sim}}
\def\OOrd{\buildrel{\scriptscriptstyle >}\over{\scriptscriptstyle\sim}}
\def\pl #1 #2 #3 {{\it Phys.~Lett.} {\bf#1} (#2) #3}
\def\np #1 #2 #3 {{\it Nucl.~Phys.} {\bf#1} (#2) #3}
\def\zp #1 #2 #3 {{\it Z.~Phys.} {\bf#1} (#2) #3}
\def\pr #1 #2 #3 {{\it Phys.~Rev.} {\bf#1} (#2) #3}
\def\prep #1 #2 #3 {{\it Phys.~Rep.} {\bf#1} (#2) #3}
\def\prl #1 #2 #3 {{\it Phys.~Rev.~Lett.} {\bf#1} (#2) #3}
\def\mpl #1 #2 #3 {{\it Mod.~Phys.~Lett.} {\bf#1} (#2) #3}
\def\rmp #1 #2 #3 {{\it Rev. Mod. Phys.} {\bf#1} (#2) #3}
\def\sjnp #1 #2 #3 {{\it Sov. J. Nucl. Phys.} {\bf#1} (#2) #3}
\def\cpc #1 #2 #3 {{\it Comp. Phys. Comm.} {\bf#1} (#2) #3}
\def\xx #1 #2 #3 {{\bf#1}, (#2) #3}
\def\preprint{{\it preprint}}

\begin{flushright}
{Cavendish-HEP-96/13}\\ 
{DFTT 48/96}\\ 
{October 1996\hspace*{.5 truecm}}\\
\end{flushright}

\vspace*{\fill}

\begin{center}
{\Large \bf 
Production of charged Higgs bosons \\
of the Minimal Supersymmetric Standard Model\\
in $b$-quark initiated processes\\
at the Large Hadron Collider}\\[1.5cm]
{\large Stefano Moretti$^{a,b}$ and Kosuke Odagiri$^{a}$}\\[0.4 cm]
{\it a) Cavendish Laboratory, University of Cambridge,}\\
{\it Madingley Road, Cambridge CB3 0HE, UK.}\\[0.25cm]
{\it b) Dipartimento di Fisica Teorica, Universit\`a di Torino,}\\
{\it and I.N.F.N., Sezione di Torino,}\\
{\it Via Pietro Giuria 1, 10125 Torino, Italy.}\\[0.5cm]
\end{center}
\vspace*{\fill}

\begin{abstract}
{\noindent 
\small We study integrated and differential rates for the
production of charged Higgs bosons $H^\pm$ of the Minimal
Supersymmetric Standard Model via $b$-quark initiated subprocesses
in $pp$ collisions at the Large Hadron Collider.
In detail, we compute cross sections and distributions of the
reactions:  
$bU\ar bDH^+\ar bD\tau^+\nu_\tau\oplus\CC$ and
$bU\ar bDH^+\ar bD t\bar b\ar bDb\bar b \mathrm{jj}\oplus\CC$, for
a $H^\pm$ scalar in the intermediate (i.e., $M^\pm<m_t+m_b$) and heavy 
(i.e., $M^\pm>m_t+m_b$) mass range, respectively ($U$ and $D$ represent generic
$u$- and $d$-type light quarks).
In the first case, a detailed treatment of various possible backgrounds is
also given.}
\end{abstract}
\vskip1.0cm
\hrule
\vskip0.25cm
\noindent
Electronic mails: moretti,odagiri@hep.phy.cam.ac.uk

\vspace*{\fill}
\newpage

\section*{1. Introduction}

At the Large Hadron Collider (LHC) \cite{CMS,ATLAS}, the charged Higgs
boson $H^\pm$ of the Minimal Supersymmetric Standard Model (\mssm)
is expected (if it exists)
to be copiously produced in top quarks decays, via the 
chain $t\ar bH^\pm \ar b(\tau\nu_\tau)$, provided that
$m_t>M_{H^\pm}+m_b$ and that the value of $\tan\beta$
is low or high enough\footnote{The minimum of the $t\ar bH^\pm$ 
decay rate is at about $\tan\beta=6$.}.  

The main decay modes of the $H^\pm$ scalar are (for $M_{H^\pm}<m_t+m_b$)
into $cs$, $\tau\nu_\tau$ and 
$W^\pm h$ pairs, where in the last case $h\ar b\bar b$ ($h$ being the
lightest neutral Higgs boson of the \mssm) \cite{ioejames}.
The first decay is never dominant, whereas the second is overcome
by the third in a narrow mass window right before the opening
of the $tb$ Higgs decay threshold, but only at very small values of 
$\tan\beta$. Otherwise, the branching ratio $\mathrm{BR}(H^\pm\ar 
\tau\nu_\tau)$ is the largest (around 
$98\%$, for $\tan\beta>2$), and it depends
only slightly on the $\beta$ angle.
When $M_{H^\pm}>m_t+m_b$, 
the $H^\pm\ar tb$ decay mode is in practise the only relevant
one (with a branching ratio of  practically 
$100\%$)\footnote{In the above discussion and throughout this paper
we have assumed that the mass scale of the Supersymmetric particles
is above the $H^\pm$ mass, such that
only decays into ordinary matter are here considered.}.

Extensive studies and simulations for $H^\pm$ production at the LHC
have been carried out \cite{CMS,ATLAS}. Top quarks
are produced in $t\bar t$ pairs, via $q\bar q$ and
$gg$ fusion, with a large cross section (around 500 pb at 
$\sqrt s_{pp}=14$ TeV and for $m_t=175$ GeV). The signal that
has been considered in the ATLAS and CMS Technical Proposals
is the one involving one top decaying to a charged Higgs, and the other
decaying inclusively (i.e., either via a $H^\pm$ or, mostly, a $W^\pm$)
into electrons and muons (and corresponding neutrinos). 
The charged Higgs boson is searched for 
by means of the leptonic signature $H^\pm\ar\tau\nu_\tau$. 

Since neutrinos in the final state prevent one from reconstructing the
Higgs mass from the momenta of its decay products, the existence
of $H^\pm$ signals in the data can be inferred only from an excess 
of $\tau$ production with respect to what is predicted for the Standard
Model (\sm) backgrounds ({\sl lepton universality breaking} signal).
Among the letter, one must number the irreducible ones (non-resonant
$\tau$ production and  
$t\bar t$ production followed by $t\ar W^\pm b$ and $W^\pm\ar \tau\nu_\tau$)
as well as the reducible ones (mainly $t\bar t$ where either a jet from a
$W^\pm$ fakes a $\tau$ or a $b$-jet 
decays leptonically into or fakes a $\tau$, 
but also $b\bar b$ production followed by $b\bar b\ar \tau$ + jets and
$W^\pm$ + jets, with one of the jets faking a $\tau$) \cite{CMS,ATLAS}.
By selecting an isolated high $p_T$ lepton and by
requiring one jet to have a high transverse energy together with 
one $b$-tagging should allow one to explore a large
portion of the $(M_A,\tan\beta)$ plane, with a  significance up to $5\sigma$ 
(assuming a 10 fb$^{-1}$ integrated luminosity of the collider) \cite{CMS}.

It is the purpose of this letter to study other production mechanisms
of charged Higgs bosons of the \mssm\ at the LHC, via subprocesses with
$b$-quarks in the initial state. In particular, we calculate
(for a $H^\pm$ scalar whose mass is below the $tb$ threshold)
the signal reaction (including charge conjugation)
\be\label{mssm}
bU\ar bDH^+\ar bD\tau^+\nu_\tau\oplus\CC,
\ee
and the background processes 
\be\label{ew}
bU\ar bD\tau^+\nu_\tau\oplus\CC,
\qquad\qquad{\mathrm{via~intermediate}}~W^\pm,\gamma,Z,H,h,A,
\ee
\be\label{qcd}
bU\ar bD\tau^+\nu_\tau\oplus\CC,
\qquad\qquad{\mathrm{via~intermediate}}~W^\pm,g,
\ee
\be\label{bbbar}
U\bar D\ar b\bar b \tau^+\nu_\tau\oplus\CC,
\qquad\qquad{\mathrm{via~intermediate}}~W^\pm,g,
\ee
where $U(D)$ represents a generic $u(d)$-type light quark (i.e., $u,d,s$ and
$c$) found inside the
proton and $H,h,A$ are the three neutral Higgs bosons of the \mssm.
For a heavy $H^\pm$, we consider the production and decay chain
(again including charge conjugation)
\be\label{mssm_new}
bU\ar bDH^+\ar bD \bar bt\ar bD\bar b b\mathrm{jj}\oplus\CC,
\ee
assuming that, because of the spectacular signature that is produced
in the final state, background processes can easily be kept under 
control. In reactions (\ref{mssm})--(3) and (\ref{mssm_new}) we treat
the initial $b$-quark as a constituent of the proton with the appropriate 
momentum fraction distribution $f_{b/p}(x,Q^2)$, as given by our default set
of partonic structure functions.

The relevance of these reactions can be understood if one considers that:
\begin{itemize}
\item at the typical (partonic) energies of the LHC 
the content of $b$-quarks inside the
colliding protons is very much enhanced, compared to lower energy hadronic
scatterings (such as, e.g., at the Tevatron);
\item the presence of $b$-quark in initial, virtual and final states
means that many of the \mssm\ Yukawa couplings of the Higgs bosons
of the theory are increased for large values of $\tan\beta$;
\item the vertex tagging performances of the LHC detectors are expected to
become almost `ideal' by the time the machine starts operations, thus
allowing one to greatly 
reduce the QCD background of light quark and gluon jets.   
\end{itemize}
Finally, we also stress that $b$-quark initiated processes at LHC energies have
already been demonstrated to be important in the case of neutral \mssm\ Higgs 
production, especially at large $\tan\beta$'s and for intermediate
masses of the scalars \cite{Pistarino}, as well as in the case of charged Higgs
production, via the reaction $g\bar b\ar H^+\bar t\oplus\CC$ \cite{bg}.

The plan of this paper is as follows. In the next Section we
give some details of the calculation
and list the values adopted for the various parameters.
Section 3 is devoted to a discussion of the results.
Conclusions are in Section 4. In the Appendix we write down the amplitudes
squared of the signal processes.

\section*{2. Calculation} 

To calculate processes (\ref{mssm})--(\ref{mssm_new}) we have used the 
spinor techniques described in Refs.~\cite{KS,mana,ioPR}. The {\tt
FORTRAN} codes we have produced have been counter-checked against the
outputs of
MadGraph \cite{tim}, which incorporates the {\tt HELAS} subroutines
\cite{HELAS}. We have always found perfect agreement
between the two kind of programs. 
The codes written using the helicity formalism of 
Refs.~\cite{KS,mana,ioPR} have also been tested for gauge invariance.
Furthermore, in order to speed up the numerical evaluations in Monte Carlo 
simulations, the matrix elements for the signal processes  
(\ref{mssm}) and (\ref{mssm_new}) have been computed 
using the textbook method of taking the trace of
 the gamma matrices, with the help of
FORM \cite{FORM}. The analytical expressions obtained in this way are
very simple, so that
we do reproduce them here (see the Appendix). They have been eventually
implemented and their numerical results agree with those of the other
codes. The integrations over the phase space have been performed using
{\tt VEGAS} \cite{VEGAS}. 

The tree-level Feynman diagrams that one needs for computing
processes (\ref{mssm})--(\ref{mssm_new}) are given in Fig.~1a,b,c
and d\footnote{The \CC\ diagrams can be obtained by simple
crossing and time inversion of fermion lines.}.
The labelling of the particles in the figures corresponds
to their ordering in the left-hand side (for the initial state) and
in the right-hand side (for the final state) of
equations (\ref{mssm})--(\ref{mssm_new}).
Note that the virtual particle content of the diagrams 
is explicitly indicated in Fig.~1 for all reactions.

Concerning the values of the various parameters entering in the
computation of processes (\ref{mssm})--(\ref{mssm_new}), we have
proceeded as follows. First, we have set up the mass scale of the
Supersymmetric partners of ordinary matter well
above the energy reach of the LHC,
such that we can neglect their contribution in our calculations.
To further  simplify the discussion, we have assumed a universal
soft Supersymmetry--breaking mass \cite{corrMH0iMSSM,corrMHMSSM}
\be
m_{\tilde u}^2=m_{\tilde d}^2=m_{\tilde q}^2,
\ee
and negligible mixing in the stop and sbottom mass matrices,
\be
A_t=A_b=\mu=0.
\ee
One-loop corrections to the masses of the ${\cal {MSSM}}$
  neutral ${\cal {CP}}$--even
Higgs bosons and to the mixing angle $\alpha$
are introduced via the
 parameter $\varepsilon$ of Ref.~\cite{0pmLEPLHCSSC}, given by (neglecting
the $b$-mass)
\begin{equation}\label{m2}
\varepsilon = \frac{3e^{2}}{8\pi^{2} M^{2}_{W^\pm}\sin^2\theta_W}\;  m_{t}^{4}
\;  {\mathrm {ln}}\left( 1 +
\frac{{m}^{2}_{\tilde q}}{m_{t}^{2}} \right),
\end{equation}
where $e^2=4\pi\alpha_{em}$.
One then gets \cite{corrMH0iMSSM}
\begin{eqnarray}\label{m1}
M^{2}_{h,H}& = & \frac{1}{2}[M^{2}_{A} + M_{Z}^{2} 
+ \varepsilon/\sin^{2}\beta] \nonumber \\
&   & \pm \left\{ [ (M^{2}_{A} 
- M^{2}_{Z})\cos2\beta + \varepsilon/\sin^{2}\beta]^{2}
+(M^{2}_{A} + M^{2}_{Z})^{2}{\mathrm {sin}}^{2}2\beta \right\}^{1/2},
\end{eqnarray}
and
\begin{equation}\label{m3}
\tan 2\alpha = 
\frac{(M_{A}^{2} + 
M_{Z}^{2}){\mathrm {sin}}2\beta}{(M_{A}^{2} - M_{Z}^{2})
{\mathrm {cos2}}\beta + \varepsilon/{\mathrm {sin}}^{2}\beta}.
\end{equation}
For the ${\cal {MSSM}}$ charged Higgs masses we have maintained the
tree--level relations
\begin{equation}
M_{H^\pm}^2=M_{A}^2+M_{W^\pm}^2,
\end{equation}
since one--loop corrections are small compared
to those for the neutral Higgses \cite{corrMHMSSM}.

In the numerical calculations presented in the next Section
we have adopted the following values for the electromagnetic coupling constant
and the weak mixing angle:
$\alpha_{em}= 1/128$ and  $\sin^2\theta_W=0.2320$.
The strong coupling constant $\alpha_s$, which appears at next-to-leading
order in the computation of the charged Higgs decay width (see
Ref.~\cite{ioejames}) and enters in some of the
production mechanisms, has been evaluated
at two loops, with $\Lambda^{(4)}_{\overline{\mathrm {MS}}}=230$ MeV, and with
the number of active flavours $N_f$ (and the corresponding
$\Lambda^{(N_f)}_{\overline{\mathrm{MS}}}$) calculated according to the
prescription of Ref.~\cite{MARCIANO} at
the scale $Q^2=s$. 

For the gauge boson masses and widths we have taken
$M_{Z}=91.1888$ GeV, $\Gamma_{Z}=2.5$ GeV,
$M_{W^\pm}=80.23$ GeV and
$\Gamma_{W^\pm}=2.08$ GeV, while for the fermion masses we have used
$m_e=m_{\nu_e}=m_{\nu_\mu}=m_{\nu_\tau}=0$,
$m_\mu=0.105$ GeV, $m_\tau=1.78$ GeV, $m_u=m_d=m_s=m_c =0$,
$m_b=4.25$ GeV
and $m_t=175$ GeV, with all widths equal to zero except
for $\Gamma_t$. We have calculated this at tree--level
within the \mssm, using the expressions
given in Refs.~\cite{widthtopMSSM,widthtopSM}\footnote{Actually, 
we have done done so only in the production processes (those represented
by the graphs in Fig.~1a,b,c and d). In fact, for process (5), to
describe the decay chain $t\ar bW^\pm\ar b\mathrm{jj}$, we have used
a Narrow Width Approximation (NWA) for the top, by 
implementing the decay formulae as given in Ref.~\cite{topdecay}. 
This has been done in order to avoid a large consume of CPU time in
computing exactly a $2\ar6$ partonic process convoluted with initial
structure functions and in presence of multiple resonant peaks in different
regions of the phase space.
We are confident that such an approximation does not
spoil the validity of our conclusions.}.
The universal Supersymmetry--breaking squark mass is
in the numerical analysis $m_{\tilde q}=1$ TeV and the LHC
centre-of-mass (CM) energy is $\sqrt s_{pp}=14$ TeV.
Finally, throughout the paper we have always used MRSA \cite{MRSA}
as the default set of partonic distributions, with the same $\alpha_s$ and
$\Lambda^{(4)}_{\overline{\mathrm{MS}}}$ as above.

\section*{3. Results}

As it is impractical to cover all possible regions of the
\mssm\ parameter space $(\MA,\tb)$, we have decided to concentrate
here on the two representative (and extreme) 
values $\tan\beta=1.5$ and 30, and on masses of the pseudoscalar Higgs
boson $A$ in the range 60 GeV $\Ord\MA\Ord$ 500 GeV.
The large bibliography existing on the \mssm\ Higgs decay phenomenology 
allows one to easily extrapolate our results to other values of \tb\
\cite{guide}. 

In Fig.~2 we display the total cross sections at the LHC for
processes (1)--(4), for values of $\MA$ up to 480 GeV and for
the two above-mentioned $\tb$'s (note that a minimum transverse
momentum of 10 GeV is required for all detectable particles in the final 
states of all processes).
The relevant feature of Fig.~2 is that process (1) has rather
large rates, of the same order as the backgrounds (2)--(4). This
is particularly true below $\MA\approx120$ GeV (which corresponds to
$\MHpm\approx145$ GeV, small window in the upper right corner of Fig.~2), 
and more for large than for small values
of $\tb$. The latter aspect is a consequence of the fact
that the BR of the charged Higgs boson into $\tau\nu_\tau$
pairs is enhanced at $\tbb$, a value for which the $cs$ channel is negligible
(see Ref.~\cite{ioejames}). An additional contribution 
comes from graph 2 in Fig.~1a, which involves Yukawa vertex contributions
proportional to $\tb$. However, the largest part of the signal cross
section is due to graph 1 in Fig.~1a, because of a resonant top decay.
The steep decrease of the signal rates around $\MA\approx 150$ GeV is 
due to the opening of the $H^\pm\ar tb$ off-shell decay channel (see 
Refs.~\cite{ioejames,them}).

Concerning the background processes (2)--(4), one notices that they
are `roughly' independent of $\MA$ and $\tb$.
This is obvious for processes (3)--(4), as they proceed through \sm\
graphs (Figs.~1c and d), whereas for process (2) this indicates that 
the contributions to the total cross section due to interactions involving
\mssm\ vertices (that is, graphs 8 and 11 in Fig.~1b) are 
irrelevant (even at $\tbb$).
We also notice that, unlike processes (2)--(3), reaction (4)
is not an irreducible background, as its final state is different
from that of the signal. Nevertheless, as it includes two bottom
quarks among the produced particles, the probability
that its final signature be the same as the signal is 
(neglecting the
misidentification of light flavour quarks as $b$'s as well
as correlations between the
two possible tags): $P=1-B$, 
where $B=2\varepsilon_{b}-\varepsilon_{b}^2$, 
$\varepsilon_{b}$ being the efficiency of tagging one $b$-jet.
Hence, better $b$-tagging leads to lower detection
rates of the background process (4). 

Therefore, from the figure it is clear that in principle
a large excess of $\tau$ events could be produced in the scattering
process, provided 
that $\MA\Ord120-130$ GeV (at large $\tb$'s such an interval can be possibly
extended up to 130--140
 GeV). The possibilities of actually disentangling the signal
depend strongly on the
detector performances of the LHC, in particular in recognising the
displaced vertex in jets which originate from 
$b$-quarks\footnote{We believe that hadronic $\tau$ decays 
can be easily distinguished from quark and gluon jets.}. In fact, the
signature that one would look for is 
$b\mathrm{j}\tau{E_T\Dirin}X$, where ${E_T\Dirin}$
represents the missing (transverse) energy due to the neutrino escaping the
detectors and j is the
jet arising from the light parton scattered in the 
proton. 
In order to quantify the significance of the signals
we list in Tab.~I the total cross sections of processes 
(1)--(4) (as read from Fig.~2) for
$\MA=60(80)[100]\{120\}$ GeV
(corresponding to 
$\MHpm=100(113)[128]\{144\}$ GeV), for both values of $\tan\beta=1.5$ and 30, 
multiplied by the $b$-tagging efficiencies and by the yearly
luminosity $\int{\cal L}dt=10~\mathrm{fb}^{-1}$ (the minimum considered
for the final collider design). Furthermore, we also compute the relative excess
of signal events $S$ respect to the total background $B$, as 
$\Delta=S/B\times100$.
For the microvertex performances, we have adopted the following three
reference values: $\varepsilon_{b}=1.0,0.75$ and 0.5\footnote{In first
approximation, we assume
that the rejection factor for misidentification of light quarks and
gluons as $b$-jets is large enough, that we can neglect 
the reducible QCD background here.}. 
From the numbers given there one deduces that the relative excess 
of $H^\pm$ events is always quite 
large, especially at $\tbb$, and that the absolute statistics 
is huge, between ${\cal O}(10^4-10^5)$ signal events per year.
In our opinion then, given the expected performances of the LHC
detectors \cite{CMS,ATLAS}, 
such signals could well be detectable soon after turning on
the machine.
Moreover, we stress that, as the total
luminosity gets larger, the significance of the signal with
respect to the total background will increase further.

However, before drawing optimistic conclusions,
one has to carefully consider first the kinematic properties
of processes (1)--(4),
as the LHC detectors will have a finite coverage (for example,
in pseudorapidity
$\eta$ and transverse momentum $p_T$ of the visible particles).
Hence, we have plotted in Figs.~3 and 4, the differential spectra in the
above variables for both signal and background processes. 
If one assumes that the phase space region that can be covered experimentally 
is approximately
the one delimited by the requirements $|\eta(b,\tau,\mathrm{j})|<3$ and 
$p_T(b,\tau,\mathrm{j})>20$ GeV
\cite{CMS,ATLAS}, then one can easily verify that most 
signal events are contained in the detectable region (we have checked
that also in the case $\MA=60$ GeV one gets distributions similar 
to those of Fig.~3 and 4). This
is true for background events as well, process (4) being possibly the
only exception (in the pseudorapidity spectra, upper left corner in Fig.~4a
and b). In the end then, one should expect that usual selection criteria
will not alter the conclusions that were previously extracted 
from the total rates of Tab.~I. 

If the charged Higgs mass is above the $tb$ threshold, 
$H^\pm$ scalars could reveal themselves via the 
production and decay mechanism (5).
For a heavy $H^\pm$ boson, we consider then the signature 
$bbb\mathrm{jjj}X$, where $b$ represents either a quark or the corresponding
antiquark, and j a jet that does not show a displaced
vertex. 
In this case, the final state that should be detected is much more
complicated than that of an intermediate mass $H^\pm$ boson, as it is
made up of six jets. 
Nevertheless, the complex
resonant structure of process (5) brings some advantages to
the tagging procedure.
In fact, on the one hand,
three vertex tags are now required
(which introduce the suppression factor $\varepsilon_b^3$), and  
in a high hadronic multiplicity environment; in addition, 
complications arise from the combinatorics of the the jets.
On the other hand, the kinematics of the jets in the final
state is highly constrained, since: (i) one of 
three possible jj combinations 
must reproduce the $W^\pm$ mass
(i.e., $M_{\mathrm{jj}}\approx M_{W^\pm}$); (ii) one of the nine
possible $b\mathrm{jj}$ combinations must reproduce at the same time
the $t$ and $W^\pm$ masses (i.e., $M_{b\mathrm{jj}}\approx m_{t}$,
with $M_{\mathrm{jj}}\approx M_{W^\pm}$). In this respect, note that
in the decay chain $t\ar bW^\pm\ar b \mathrm{jj}$ we
do not need to consider intermediate $H^\pm$
contributions, as in the heavy $\MHpm$ range the $t\ar bH^\pm$
is forbidden. 

The total cross section for process (5)
is displayed in Fig.~5, as a function of the $A$ mass
 in the range 140 GeV $\Ord\MA\Ord$ 480 GeV (for both $\tb=1.5$
and 30). Both
the $t\ar bW^\pm$ and $W^\pm\ar \mathrm{jj}$ branching ratios are included.
For comparison, we also reproduce from Fig.~1 the rates for process
(1).
A first feature that is worth noticing in Fig.~5 is that again the rates of 
process (5) for $\tbb$ are larger than those for $\tba$.
In practice, there are two opposite effects which take place: 
on the one hand, at small $\tb$, the $BR(H^\pm\ar tb)$ is larger
whereas, on the other hand, at large $\tb$, the contribution from
graph 2 in Fig.~1a is enhanced by couplings proportional to 
$\tb$ itself\footnote{Also note that at the same time, since $\MHpm>m_t+m_b$, 
the first graph in Fig.~1a is no longer resonant.}.
Of the two, it is the second that dominates over most of
the $\MHpm$ range. Furthermore, at large $\tb$'s, processes (1) and (5)
yield rates of the same order for $\MA\OOrd200$ GeV, such that in this
case they can be contemporaneously exploited in searching for $H^\pm$ signals.
This is not true at small $\tb$.  
In general, it should be noted that in Fig.~5
one is dealing with total rates that are more than two
orders of magnitude smaller that for the case of an intermediate
mass $H^\pm$ boson in the $\tau\nu_\tau$ channel.
Nevertheless, as a starting point one can rely on ${\cal O}(10-100)$
signal events produced via $H^\pm\ar tb$
per year (for the same luminosity as above).
Clearly, in this case the need for a high value of $\varepsilon_b$
is crucial. For example, a vertex tagging efficiency of $50\%$
reduces the production rates by a factor of 8. 

In Fig.~6 we plot the spectrum in the sum of the
invariant mass of all possible
$bb\mathrm{jj}$ combinations
entering in process (5)\footnote{Note that in our NWA approach
we automatically obtain $M_{b\mathrm{jj}}\equiv m_{t}$ for the right
three jet combination. In the case of the $W^\pm$ decay we have adopted 
the `conservative' requirement $|M_{\mathrm{jj}}- M_{W^\pm}|<15$ GeV.}.
The values of $\MA$ considered here as a reference are 200, 300
400 and 500 GeV. Since at least one of the $bb\mathrm{jj}$ systems is made
up by the
decay products of the charged Higgs boson, a peak should possibly appear
in the distributions (at $\MHpm=215,311,408$ and
506 GeV, respectively), on top of the combinatorial background.
Indeed, the resonant peaks are quite sharp (the $H^\pm$ widths
are approximately, 
for the above values of mass: 0.84, 3.03, 5.20 and 7.20 GeV
at both $\tb$ values), and clearly
visible. However, the total number of events in the region, say, 
$|M_{bb\mathrm{jj}}-M_{H^\pm}|<25$ GeV is, for $\tbb$: 
$25(10)[3]$, $9(4)[1]$, $4(2)[0.4]$ and $2(1)[0.2]$   
(for the four masses above), for $10~\mathrm{fb}^{-1}$ per year of luminosity,
assuming $\varepsilon_b=1(0.75)[0.5]$. Therefore, at least for
not too heavy $H^\pm$'s and high $b$-tagging performances, one
could possibly look for $H^\pm$ signals in the $tb$ channel at large $\tb$'s.
Certainly, if the high luminosity option $\int{\cal L}dt=100~\mathrm{fb}^{-1}$ 
can be achieved at the LHC, things would be very optimistic, as in a few
years of running, even the very heavy mass region could be scanned. 
At small values of $\tb$ one has to consider rates that are typically smaller
by one order of magnitude, rendering Higgs detection much more difficult.

Finally, in Fig.~7, we show the differential spectrum
in transverse momentum of the various $bb\mathrm{jj}$ combinations
that can be reconstructed from process (5), for the same values
of \MA\ as above. We notice that from the figure it is
clear that the $p_T$ spectrum of the Higgs decay products
is significantly hard (because of the large
mass of the scalar), a feature that could well help 
in disentangling
heavy $H^\pm$ boson signals, especially considering that the ordinary QCD 
background in six-jet events has quite a soft transverse momentum
distribution.

\section*{4. Conclusions}

In this paper we have studied, within the \mssm, a new production
mechanism of charged Higgs bosons $H^\pm$  at LHC energies, 
via $b$-quark initiated interactions. Two possible Higgs
signatures have been considered,
depending on whether the mass $\MHpm$ is below or above 
$m_t+m_b\approx180$ GeV. 

In the first case, by exploiting one vertex tag on one
$b$-jet in the final state, the Higgs decay channel 
$H^\pm\ar\tau\nu_\tau$ should be 
identifiable as a clear excess in the number 
of $\tau$ events with respect to the rates predicted by the non-SUSY 
backgrounds, provided that $\MA\Ord130$ GeV (i.e.,
$\MHpm\Ord144$ GeV), both at large and small values of 
$\tan\beta$. Also, the absolute number of signal
events is statistically very large.
In this mass range,  
a careful treatment of various background sources
has been performed.

In the second case, the channel
$H^\pm\ar tb\ar bb\mathrm{jj}$ (via hadronic $W^\pm$ decays of the top) has
been considered. The signature arising from heavy $H^\pm$ decays
is complicated (a six-jet final state involving three $b$-jets)
and only a small number of events is expected.
Nonetheless, the
resonant behaviour of the $t$, $H^\pm$ and $W^\pm$ decay products
should allow one to eliminate the ordinary QCD background in light
quark and gluon jets, although (in the heavy mass range) a signal-to-background
analysis has not been performed. In general, if high $b$-tagging performances
can be achieved and/or the high luminosity option becomes available
at the LHC, $H^\pm$ scalars with masses up to 500 GeV could well be
searched for, as the combinatorial background does not spoil the
form of the Higgs peaks. This is however true only for large values of \tb,
since for low $\tb$'s the event rates are smaller by one order of magnitude.

The range between $\MA\approx130-140$ GeV and up to the opening 
of the $tb$ decay threshold is extremely difficult to cover, as rates
in the $\tau\nu_\tau$ channel drastically decrease well below the background
rates and at the same time the off-shell $tb$ channel has a very small 
statistics.  

Finally, we stress that, 
before drawing any firm conclusion from our results, one should include
a realistic simulation of the expected performances of the LHC detectors
and that, in the heavy mass range $\MHpm>m_t+m_b$, a detailed 
background study (including all the hadronisation effects in a six-jet
final state, an analysis which was beyond our capabilities) should be 
performed. Nevertheless, we believe that the matter
presented here would deserve experimental attention when proceeding
to the various
simulations of the \mssm\ Higgs phenomenology at the CERN hadron collider.

\subsection*{Acknowledgements}

We thank Gavin Salam for reading the preliminary version of the 
present manuscript.
This work is supported in part by the
Ministero dell' Universit\`a e della Ricerca Scientifica, the UK PPARC,
and   the EC Programme
``Human Capital and Mobility'', Network ``Physics at High Energy
Colliders'', contract CHRX-CT93-0357, DG 12 COMA (SM).
KO is grateful to Trinity College and the Committee of Vice-Chancellors and 
Principals of the Universities of the United Kingdom for
financial support.

\subsection*{Appendix}

In this additional section we write down in analytic form 
the matrix element for the signal processes. As an example, we reproduce
that of the reaction $bU\ar bDH^+\ar bD\tau^+\nu_\tau\oplus\CC$. However, 
by replacing $\tau\ar t$ and $\nu_\tau\ar b$ one can easily obtain that for
top-bottom (on-shell) production. In fact, we have used  a 
Narrow Width Approximation for the top when this is produced from the 
$H^\pm$ splitting, by computing the exact
amplitude squared for $bU\ar bDH^+\ar bD t\bar b \oplus\CC$ and by 
interfacing this with a {\tt SUBROUTINE}  
implementing the top decay formulae into $bW^\pm$ pairs 
as given in Ref.~\cite{topdecay}. 

The matrix element squared (summed/averaged over the final/initial spins and 
colours) for the process $bU\rightarrow bDH^+\ar bD \tau^+\nu_\tau$ reads as: 
\[
|{\cal M}|^2 = |{\cal M}_0|^2 |P_H|^2 |P_W|^2 
(|{\cal M}_t|^2+|{\cal M}_\phi|^2+2 |{\cal M}_t^*{\cal M}_\phi|),
\]
with
\[
|{\cal M}_0|^2 = (g^4/2 M_W^2)^2 
[p_\tau\cdot p_{\nu_\tau}(m_\tau^2\tan^2\beta + 
m_{\nu_\tau}^2\cot^2\beta) - 2m_\tau^2 m_{\nu_\tau}^2]
\]
\[
P_H=(p_H^2-M_H^2+iM_H\Gamma_H)^{-1}
\]
\[
P_W=(p_W^2-M_W^2+iM_W\Gamma_W)^{-1}
\]
\[
|{\cal M}_t|^2=2 |P_t|^2\:p_{b,in}\cdot p_U\times
\]
\[
\times
[2m_b^2(m_t^2+p_{b,out}\cdot p_t\:\tan^2\beta)\:p_t\cdot p_D
+(m_t^4\cot^2\beta-m_b^2p_t^2\tan^2\beta)\:p_{b,out}\cdot p_D]
\]
\[
|{\cal M}_\phi|^2=m_b^2\sec^2\beta
(2p_H\cdot p_D\:p_H\cdot p_U-p_H^2\:p_U\cdot p_D)\times
\]
\[
\times
[m_b^2(|P_{H_0,h_0}|^2-|P_{A_0}|^2) + p_{b,out}\cdot p_{b,in}\:
(|P_{H_0,h_0}| ^ 2+|P_{A_0}|^2)]
\]
\[
2 |{\cal M}_t^*{\cal M}_\phi|=m_b^2\sec\beta\times
\]
\[
\times[
[\Re[P_t^*(P_{H_0,h_0}-P_{A_0})]m_t^2\cot\beta-
\Re[P_t^*(P_{H_0,h_0}+P_{A_0})]m_b^2\tan\beta]\times
\]
\[
\times
(p_{b,out}\cdot p_D\:p_H\cdot p_U+p_{b,out}\cdot p_U\:p_H\cdot p_D
-p_{b,out}\cdot p_H\:p_U\cdot p_D)+
\]
\[
+[\Re[P_t^*(P_{H_0,h_0}-P_{A_0})]m_b^2\tan\beta+
\Re[P_t^*(P_{H_0,h_0}+P_{A_0})]
(m_t^2\cot\beta+2p_{b,in}\cdot p_{b,out}\tan\beta)]\times
\]
\[
\times
(p_{b,in}\cdot p_D\:p_H\cdot p_U+p_{b,in}\cdot p_U\:p_H\cdot p_D
-p_{b,in}\cdot p_H\:p_U\cdot p_D)+
\]
\[
2\Re[P_t^*(P_{H_0,h_0}+P_{A_0})]\tan\beta\times
\]
\[
\times
(p_{b,in}\cdot p_U\:p_{b,out}\cdot p_H\:p_U\cdot p_D
-p_{b,in}\cdot p_U\:p_{b,out}\cdot p_D\:p_U\cdot p_H+
\]
\[
+p_{b,in}\cdot p_D\:p_U\cdot p_{b,out}\:p_U\cdot p_H-
p_{b,in}\cdot p_H\:p_U\cdot p_{b,out}\:p_U\cdot p_D)-
2\epsilon_{\mu\nu\lambda\sigma}
p^\mu_{b,in}p^\nu_{b,out}p^\lambda_Up^\sigma_D\times
\]
\[
\times[\Im(P_t^*P_{A_0})m_b^2\tan\beta-
\Im(P_t^*P_{H_0,h_0})m_t^2\cot\beta
\]
\[
-\Im[P_t^*(P_{H_0,h_0}+P_{A_0})]\tan\beta(p_{b,in}\cdot p_{b,out}-
p_{b,in}\cdot p_U+p_{b,out}\cdot p_U)].
\]
Here we have set $m_U=m_D=0$, $p_t=p_{b,in}+p_W=p_{b,out}+p_H$, and
\[
P_t=(p_t^2-m_t^2+im_t\Gamma_t)^{-1}
\]
\[
P_{H_0,h_0}=
\cos\alpha\sin(\beta-\alpha)(p_\phi^2-m^2_{H_0}+iM_{H_0}\Gamma_{H_0})^{-1}
+\]
\[+
\sin\alpha\cos(\beta-\alpha)(p_\phi^2-m^2_{h_0}+iM_{h_0}\Gamma_{h_0})^{-1}
\]
\[
P_{A_0}=
\sin\beta(p_\phi^2-m^2_{A_0}+iM_{A_0}\Gamma_{A_0})^{-1}
\]
\[
p_H=p_\tau+p_{\nu_\tau}\qquad\qquad p_\phi=p_{b,in}-p_{b,out} 
\qquad\qquad p_W=p_U-p_D,
\]
where $p_{b,in},p_U,p_{b,out},p_D,p_\tau$ and $p_{\nu_\tau}$ are the external 
momenta (incoming/outgoing in the initial/final state). 
Note that the symbols $\Re$ and $\Im$ refer to the real and imaginary part
of a complex number, respectively, and that $\epsilon_{\mu\nu\lambda\sigma}$
is the Levi-Civita tensor ($\epsilon_{0123}=1$).
Finally, by exchanging $p_{b,in} \leftrightarrow -p_{b,out}$ and 
$\epsilon_{\mu\nu\lambda\sigma}\leftrightarrow -\epsilon_{\mu\nu\lambda\sigma}$
one can obtain the amplitude squared for $\bar b U$ fusion,
and by relabelling $U\leftrightarrow D$ those for the $b D$-  and
$\bar b D$-initiated reactions.

\goodbreak

\section*{Table Captions}

\begin{itemize}

\item[{[I]}] Number of events for signal (1) and total background (2)--(4),
at the LHC with CM energy $\sqrt s_{pp}=14$ TeV, for
$\MA=60(80)[100]\{120\}$ GeV and $\tan\beta=1.5$ and 30,
assuming
the yearly luminosity $\int{\cal L}dt=10~\mathrm{fb}^{-1}$ and the $b$-tagging
efficiencies $\varepsilon_{b}=1.0,0.75$ and 0.5 (first, second and third row,
respectively), together with the relative excess of signal events respect
to the total background.

\end{itemize}

\section*{Figure Captions}

\begin{itemize}

\item[{[1]}] Feynman diagrams contributing at lowest order
to the subprocess $bU\ar bD\tau^+\nu_\tau\oplus\CC$:
(a) via $H^\pm$ resonant graphs;
(b) via $W^\pm$ resonant graphs and through EW only;
(c) via $W^\pm$ resonant graphs and through QCD interactions;
and to the subprocess $U\bar D\ar b\bar b\tau^+\nu_\tau\oplus\CC$:
(d) via $W^\pm$ resonant graphs and through QCD interactions.
Here, $U$ and $D$ represent any of the light quarks $u,d,s$ and $c$.
The PostScript version of the Feynman graphs has been produced using
MadGraph \cite{tim}. 

\item[{[2]}] Cross section of the four processes (1)--(4), at the LHC
for $\sqrt s_{pp}=14$ TeV, in the region
$60~\mathrm{GeV}\Ord M_A \Ord 480~\mathrm{GeV}$
and for two different values of $\tan\beta$
(in the case of processes (1) and (2)).
In the top left hand plot we enlarge the region 
60 GeV $\Ord M_A \Ord$ 150 GeV.
Solid line: process (1) with $\tan\beta=1.5$.
Dashed line (large spacing): process (1) with $\tan\beta=30$.
Dotted line (large spacing): process (2) with $\tan\beta=1.5$.
Dot-dashed line: process (2) with $\tan\beta=30$.
Dashed line (small spacing): process (3).
Dotted line (small spacing): process (4).
The cut $p_{\mathrm{T}}^{\mathrm{final}}>10$ GeV has been applied
to all particles in the final states, except for neutrinos.
The structure function set MRSA has been used.

\item[{[3]}] Differential distribution in transverse
momentum of the $b$-quark (a) and of the $\tau$-lepton (b)
for the four processes (1)--(4), at the LHC
for $\sqrt s_{pp}=14$ TeV, for the following selection of
masses $M_A=80,100,120~\mathrm{GeV}$ 
and for two different values of $\tan\beta$
(in the case of processes (1) and (2)).
Solid line: process (1) with $\tan\beta=1.5$.
Dashed line (large spacing): process (1) with $\tan\beta=30$.
Dotted line (large spacing): process (2) with $\tan\beta=1.5$.
Dot-dashed line: process (2) with $\tan\beta=30$.
Dashed line (small spacing): process (3).
Dotted line (small spacing): process (4).
Please note that the two curves corresponding to process (2) are
practically indistinguishable. 
The cut $p_{\mathrm{T}}^{\mathrm{final}}>10$ GeV has been applied
to all particles in the final states, except for neutrinos.
The structure function set MRSA has been used.

\item[{[4]}] Differential distribution in pseudorapidity
of the $b$-quark (a) and of the $\tau$-lepton (b)
for the four processes (1)--(4), at the LHC
for $\sqrt s_{pp}=14$ TeV, for the following selection of
masses $M_A=80,100,120~\mathrm{GeV}$ 
and for two different values of $\tan\beta$
(in the case of processes (1) and (2)).
Solid line: process (1) with $\tan\beta=1.5$.
Dashed line (large spacing): process (1) with $\tan\beta=30$.
Dotted line (large spacing): process (2) with $\tan\beta=1.5$.
Dot-dashed line: process (2) with $\tan\beta=30$.
Dashed line (small spacing): process (3).
Dotted line (small spacing): process (4).
Please note that the two curves corresponding to process (2) are
practically indistinguishable. 
The cut $p_{\mathrm{T}}^{\mathrm{final}}>10$ GeV has been applied
to all particles in the final states, except for neutrinos.
The structure function set MRSA has been used.

\item[{[5]}] Cross section of processes (1) and (5), at the LHC
for $\sqrt s_{pp}=14$ TeV, in the region 
$140~\mathrm{GeV}\Ord M_A \Ord 480~\mathrm{GeV}$
and for two different values of $\tan\beta$.
Solid line: process (1) with $\tan\beta=1.5$.
Dashed line (large spacing): process (1) with $\tan\beta=30$.
Dashed line (small spacing): process (5) with $\tan\beta=1.5$.
Dotted line: process (5) with $\tan\beta=30$.
The cut $p_{\mathrm{T}}^{\mathrm{final}}>10$ GeV has been applied
to all particles in the final states, including the top decay
products.
The structure function set MRSA has been used.

\item[{[6]}] Differential distribution in invariant mass
of the $bb\mathrm{jj}$ systems 
for process (5), at the LHC
for $\sqrt s_{pp}=14$ TeV, for the following selection of
masses $M_A=200,300,400,500~\mathrm{GeV}$, 
for $\tan\beta=1.5$ (solid line) and $30$ (dashed line).
The cut $p_{\mathrm{T}}^{\mathrm{final}}>10$ GeV has been applied
to all particles in the final states, including the top decay
products. Bins are 5 GeV wide.
The structure function set MRSA has been used.

\item[{[7]}] Differential distribution in transverse momentum
of the $bbjj$ systems 
for process (5), at the LHC
for $\sqrt s_{pp}=14$ TeV, for the following selection of
masses $M_A=200,300,400,500~\mathrm{GeV}$, 
for $\tan\beta=1.5$ (solid line) and $30$ (dashed line).
The cut $p_{\mathrm{T}}^{\mathrm{final}}>10$ GeV has been applied
to all particles in the final states, including the top decay
products. 
The structure function set MRSA has been used.

\end{itemize}
\vfill
\clearpage

\begin{table}%[p]%[htbp]
\begin{center}
\begin{tabular}{|c|c|c|}
\hline
\multicolumn{3}{|c|}
{\rule[0cm]{0cm}{0cm}
$N_{\mathrm{ev}}(bj\tau{E_T\Dirin}X)$}
 \\ \hline\hline
\rule[0cm]{0cm}{0cm}
$S$ & $B$ & $\Delta$ \\ \hline\hline
\rule[0cm]{0cm}{0cm}
$386(289)[170]\{62\}\times10^3$ &
$630(645)[654]\{671\}\times10^3$  &
$61(45)[26]\{9\}\%$  \\ 
$289(217)[127]\{47\}\times10^3$  &
$488(498)[506]\{518\}\times10^3$ &
$59(44)[25]\{9\}\%$ \\ 
$193(145)[85]\{31\}\times10^3$ &
$375(383)[387]\{396\}\times10^3$ &
$51(38)[22]\{8\}\%$ \\ \hline
\multicolumn{3}{|c|}
{\rule[0cm]{0cm}{0cm}
$\tba$}
\\ \hline\hline
\rule[0cm]{0cm}{0cm}
$551(436)[294]\{145\}\times10^3$  &
$623(639)[651]\{672\}\times10^3$  &
$88(68)[45]\{22\}\%$  \\ 
$413(327)[220]\{109\}\times10^3$  &
$483(494)[504]\{519\}\times10^3$  &
$86(66)[44]\{21\}\%$  \\ 
$275(218)[147]\{73\}\times10^3$   &
$372(380)[386]\{396\}\times10^3$  &
$74(57)[38]\{18\}\%$  \\ \hline
\multicolumn{3}{|c|}
{\rule[0cm]{0cm}{0cm}
$\tbb$}
\\ \hline\hline
\multicolumn{3}{|c|}
{\rule[0cm]{0cm}{0cm}
\quad\quad\quad\quad$\MA=60(80)[100]\{120\}$ GeV\quad\quad\quad\quad}
\\ \hline \hline
\multicolumn{3}{|c|}
{\rule[0cm]{0cm}{0cm}
$p_{\mathrm{T}}^{\mathrm{final}}>10$ GeV 
\quad\quad\quad\quad\quad\quad\quad\quad\quad{MRSA}}
\\ \hline \hline
\multicolumn{3}{|c|}
{\rule[0cm]{0cm}{0cm}
$\sqrt s_{pp}=14$ TeV
\quad\quad\quad\quad\quad\quad\quad\quad$\int{\cal L}dt=10$ fb$^{-1}$}
\\ \hline
\multicolumn{3}{c}
{\rule{0cm}{1cm}
{\Large Tab. I}}  \\
\multicolumn{3}{c}
{\rule{0cm}{0cm}}
\end{tabular}
\end{center}
\end{table}

\begin{figure}[p]
~\epsfig{file=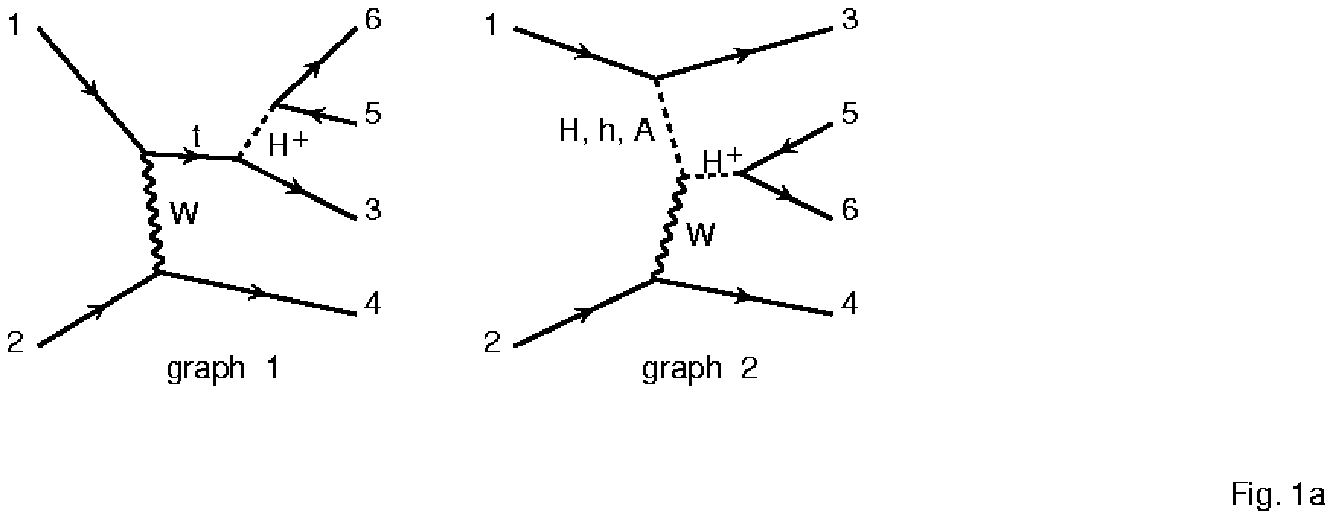,height=22cm}
\vspace*{2cm}
\end{figure}
\vfill
\clearpage

\begin{figure}[p]
~\epsfig{file=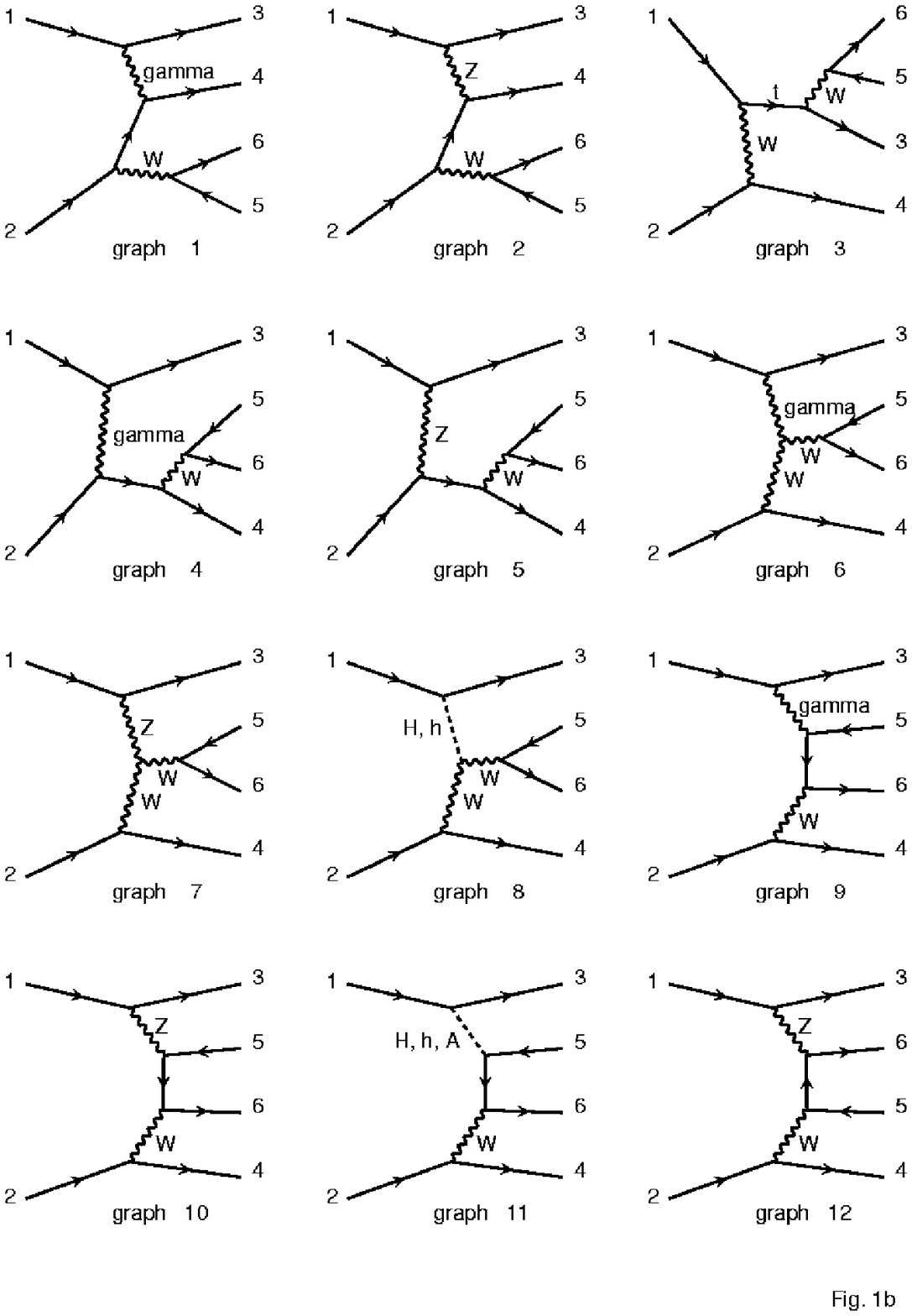,height=22cm}  
\vspace*{2cm}
\end{figure}
\vfill
\clearpage
\begin{figure}[p]
~\epsfig{file=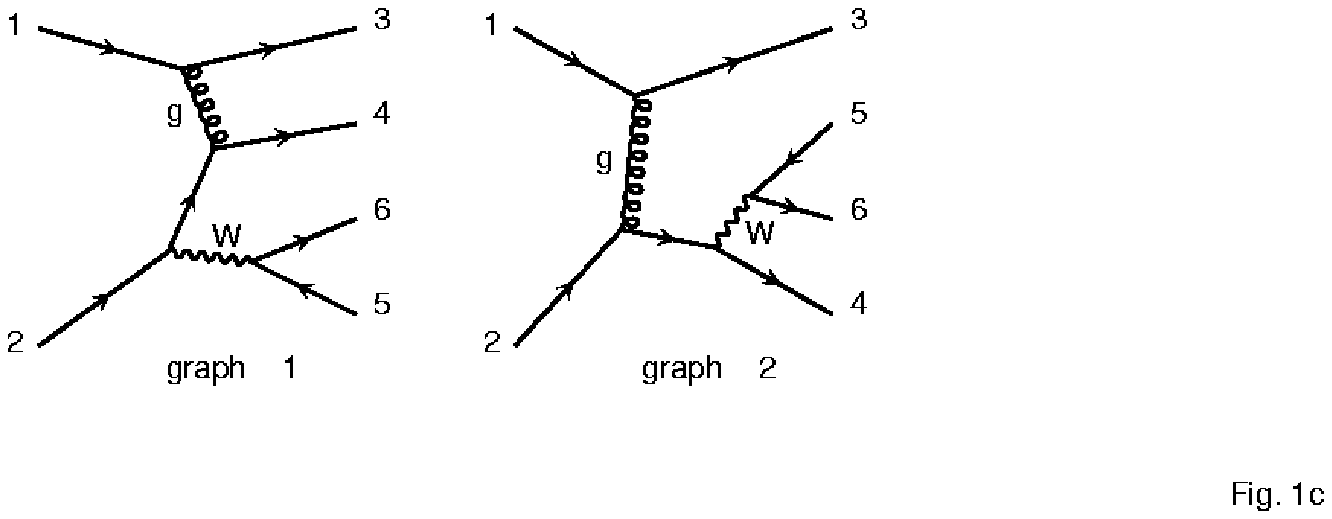,height=22cm}  
\vspace*{2cm}
\end{figure}
\vfill
\clearpage
\begin{figure}[p]
~\epsfig{file=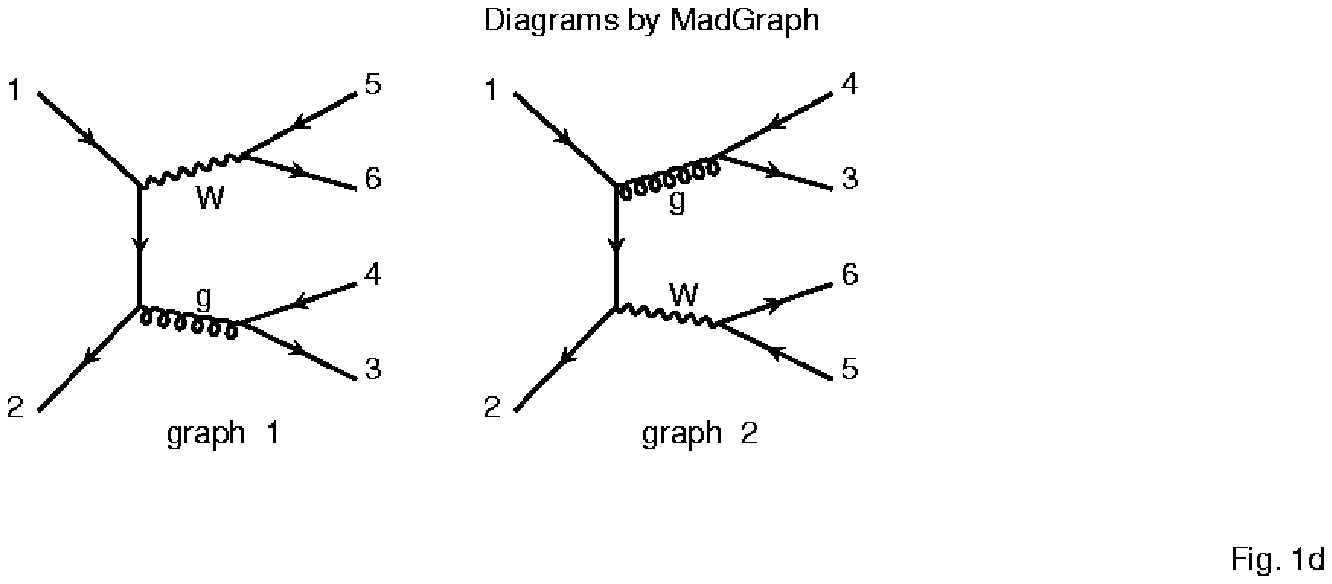,height=22cm}  
\vspace*{2cm}
\end{figure}
\vfill
\clearpage
\begin{figure}[p]
~\epsfig{file=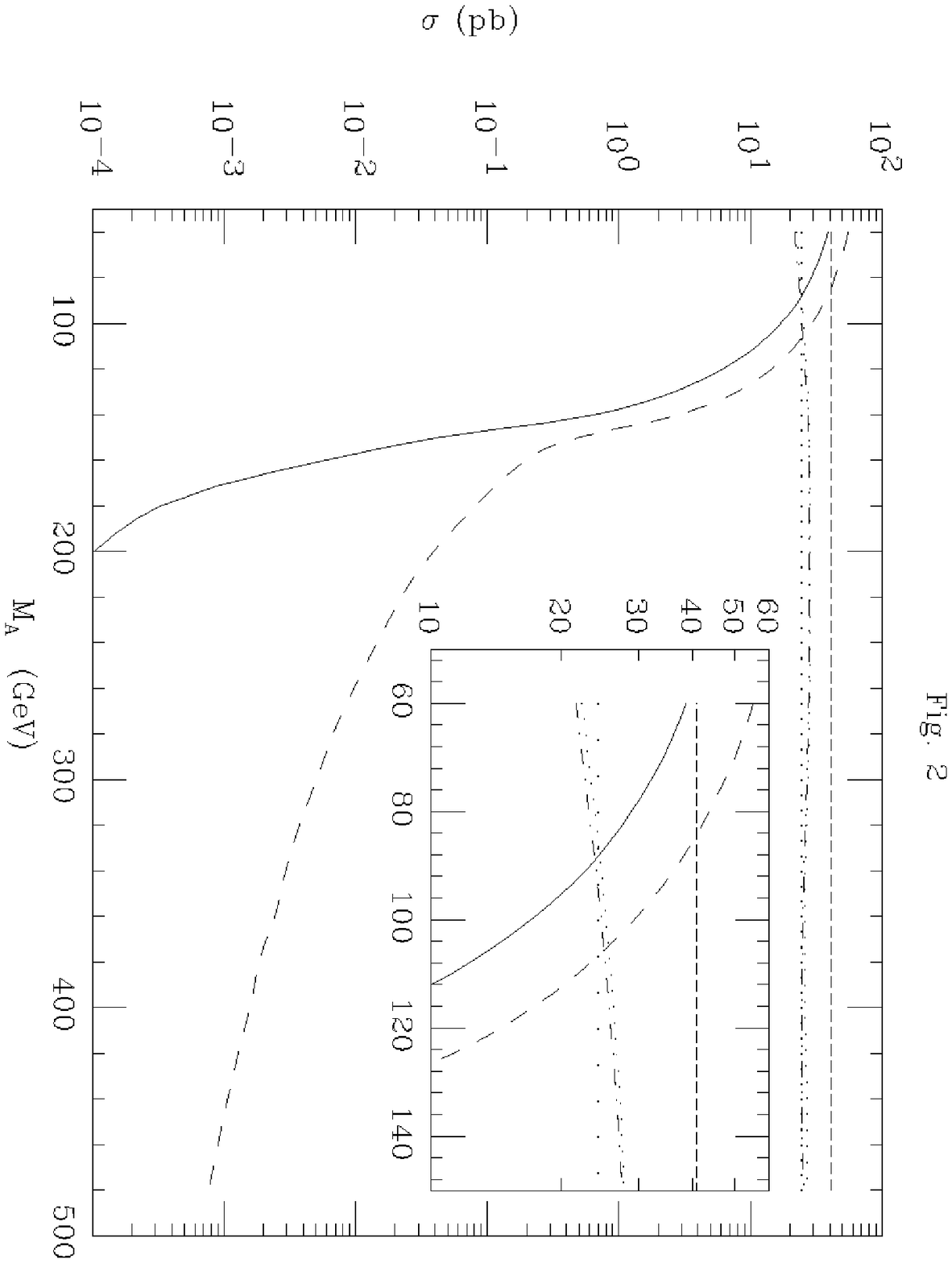,height=22cm}  
\vspace*{2cm}
\end{figure}
\clearpage
\begin{figure}[p]
~\epsfig{file=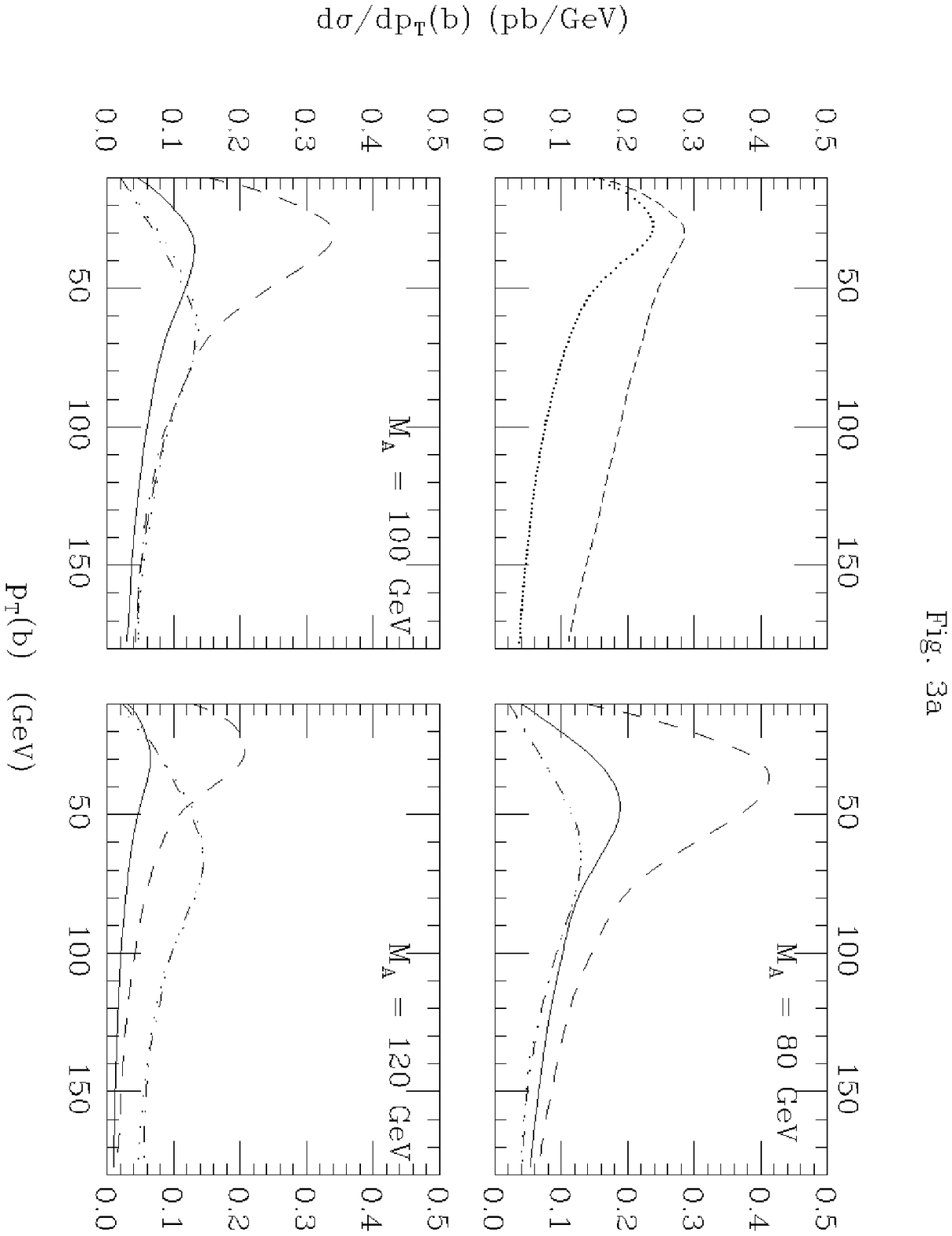,height=22cm}  
\vspace*{2cm}
\end{figure}
\clearpage
\begin{figure}[p]
~\epsfig{file=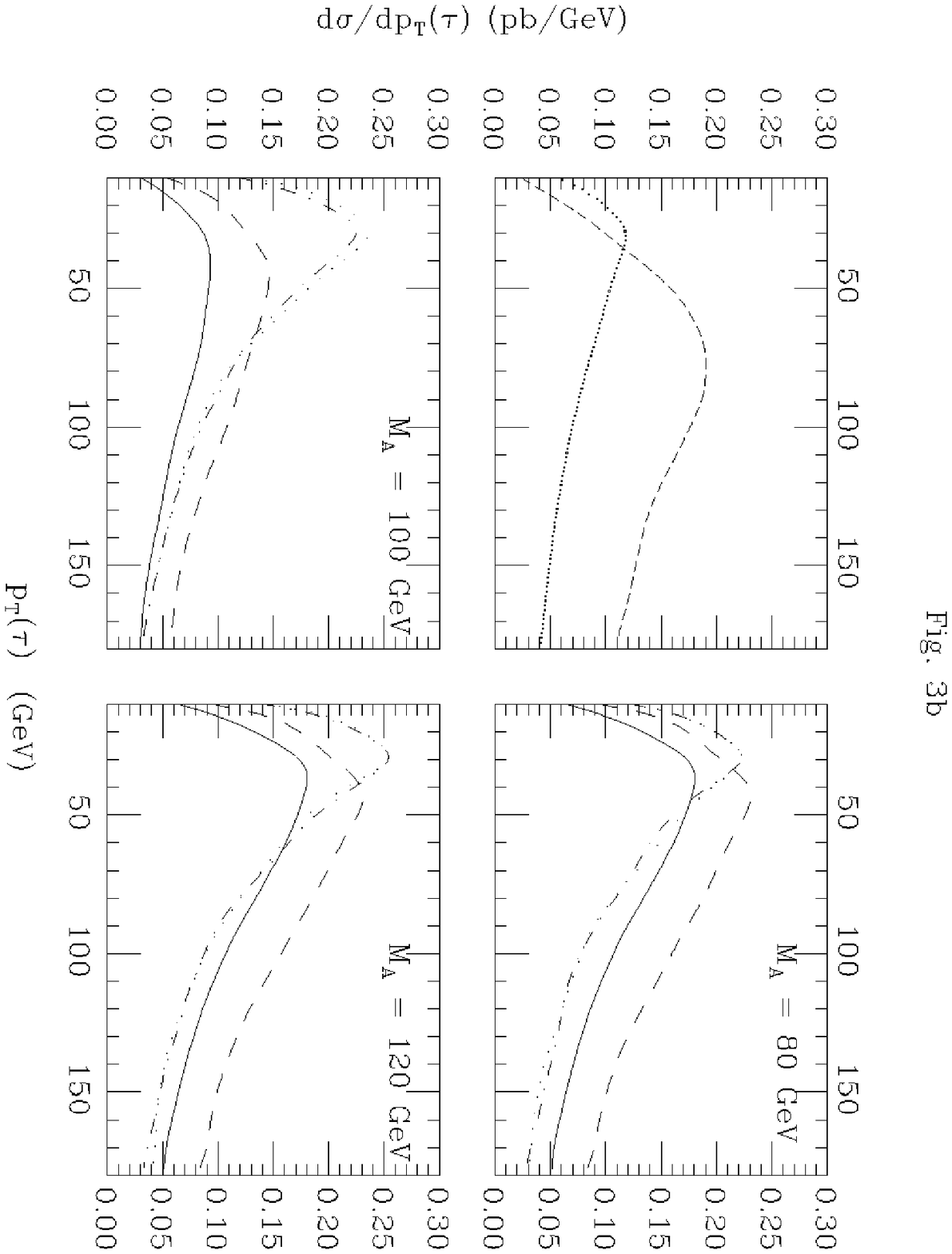,height=22cm}  
\vspace*{2cm}
\end{figure}
\clearpage
\begin{figure}[p]
~\epsfig{file=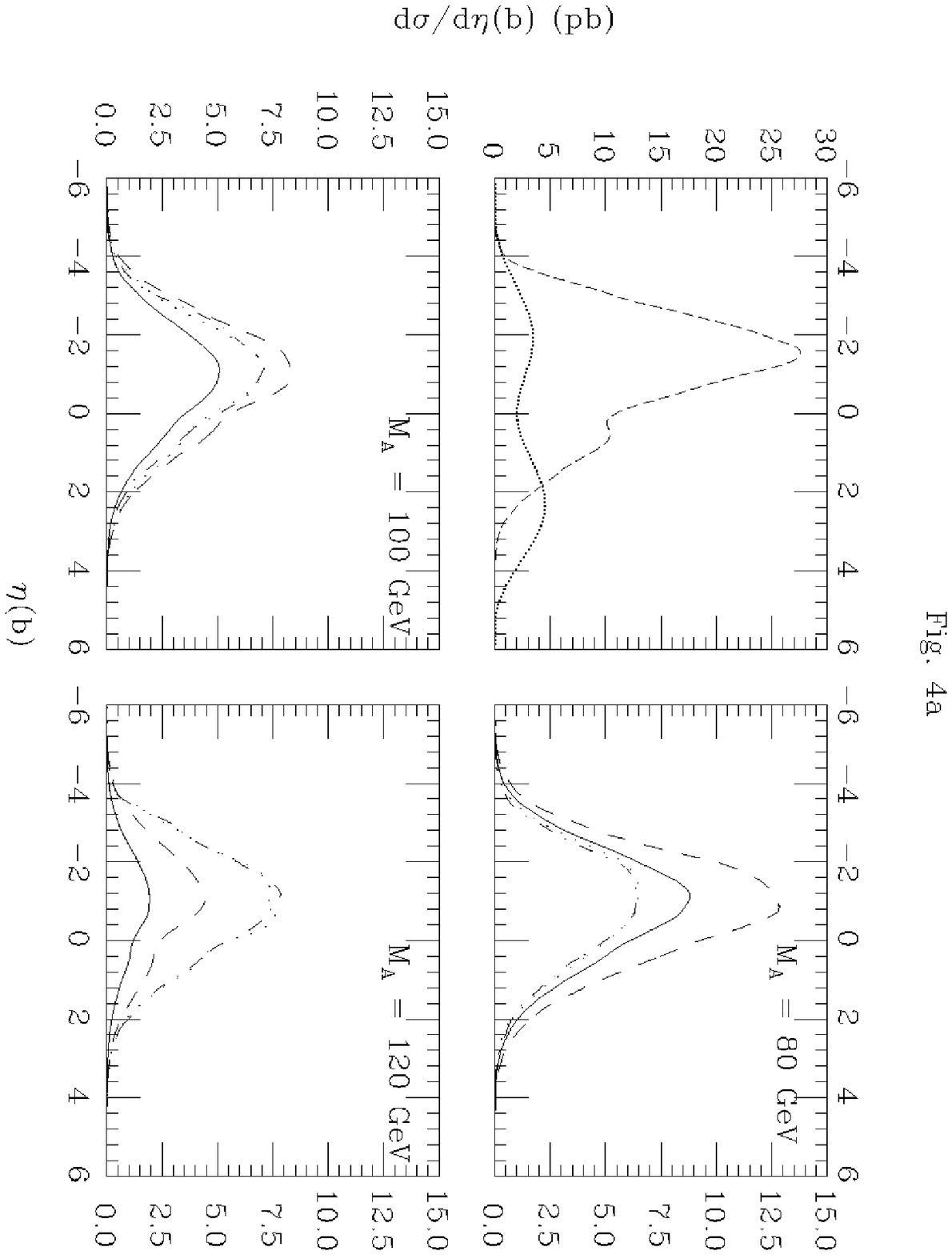,height=22cm}  
\vspace*{2cm}
\end{figure}
\clearpage
\begin{figure}[p]
~\epsfig{file=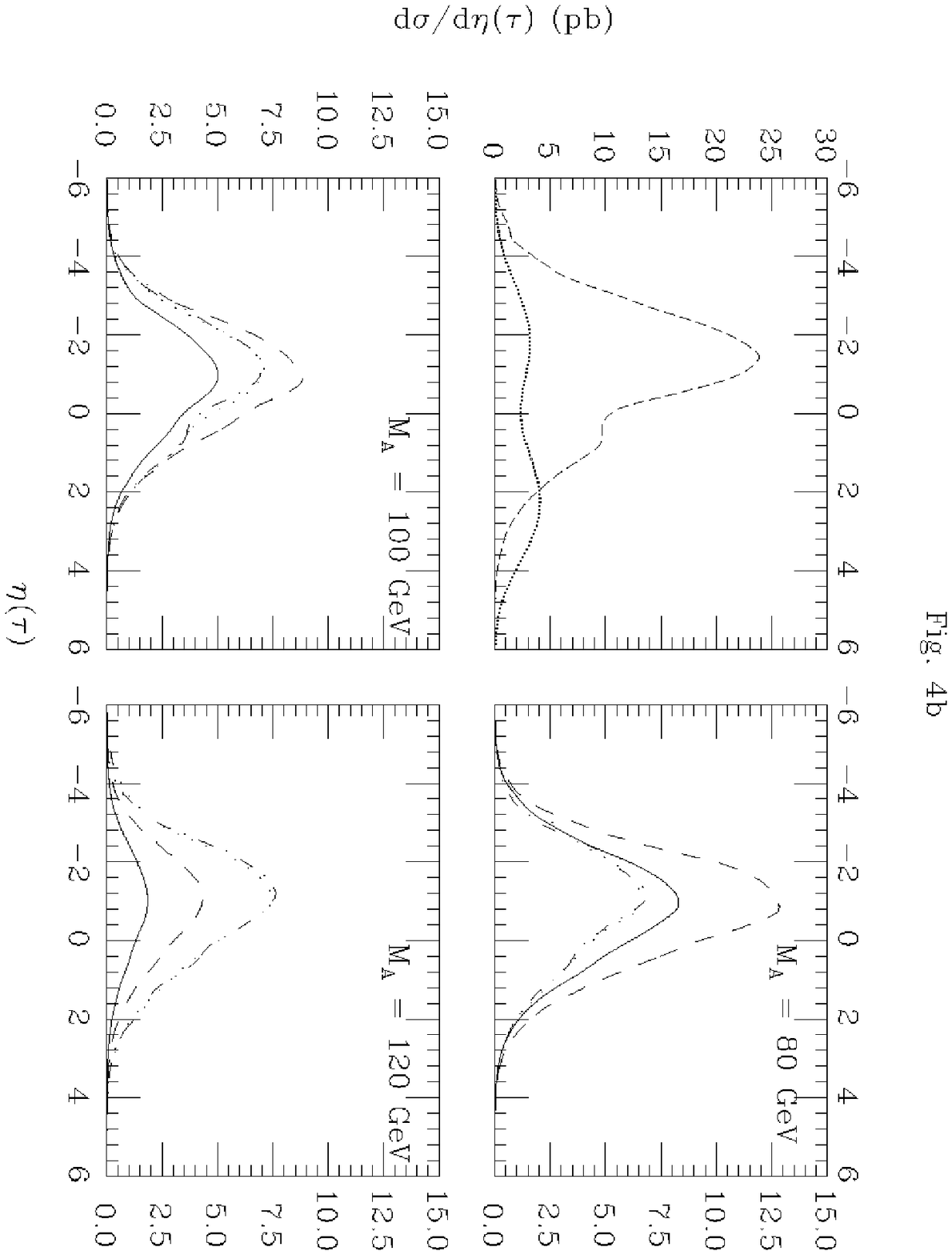,height=22cm}  
\vspace*{2cm}
\end{figure}
\clearpage
\begin{figure}[p]
~\epsfig{file=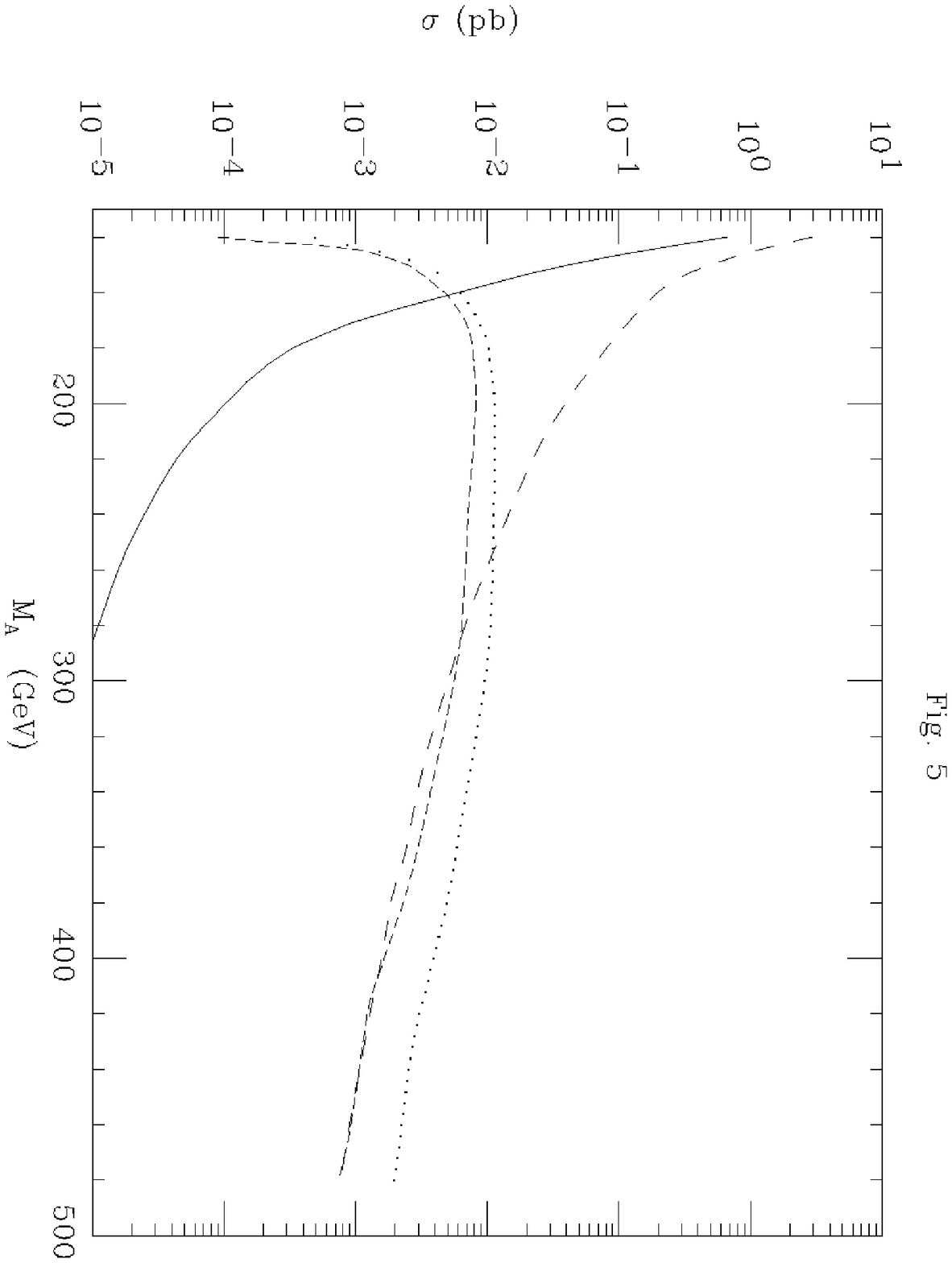,height=22cm}  
\vspace*{2cm}
\end{figure}
\clearpage
\begin{figure}[p]
~\epsfig{file=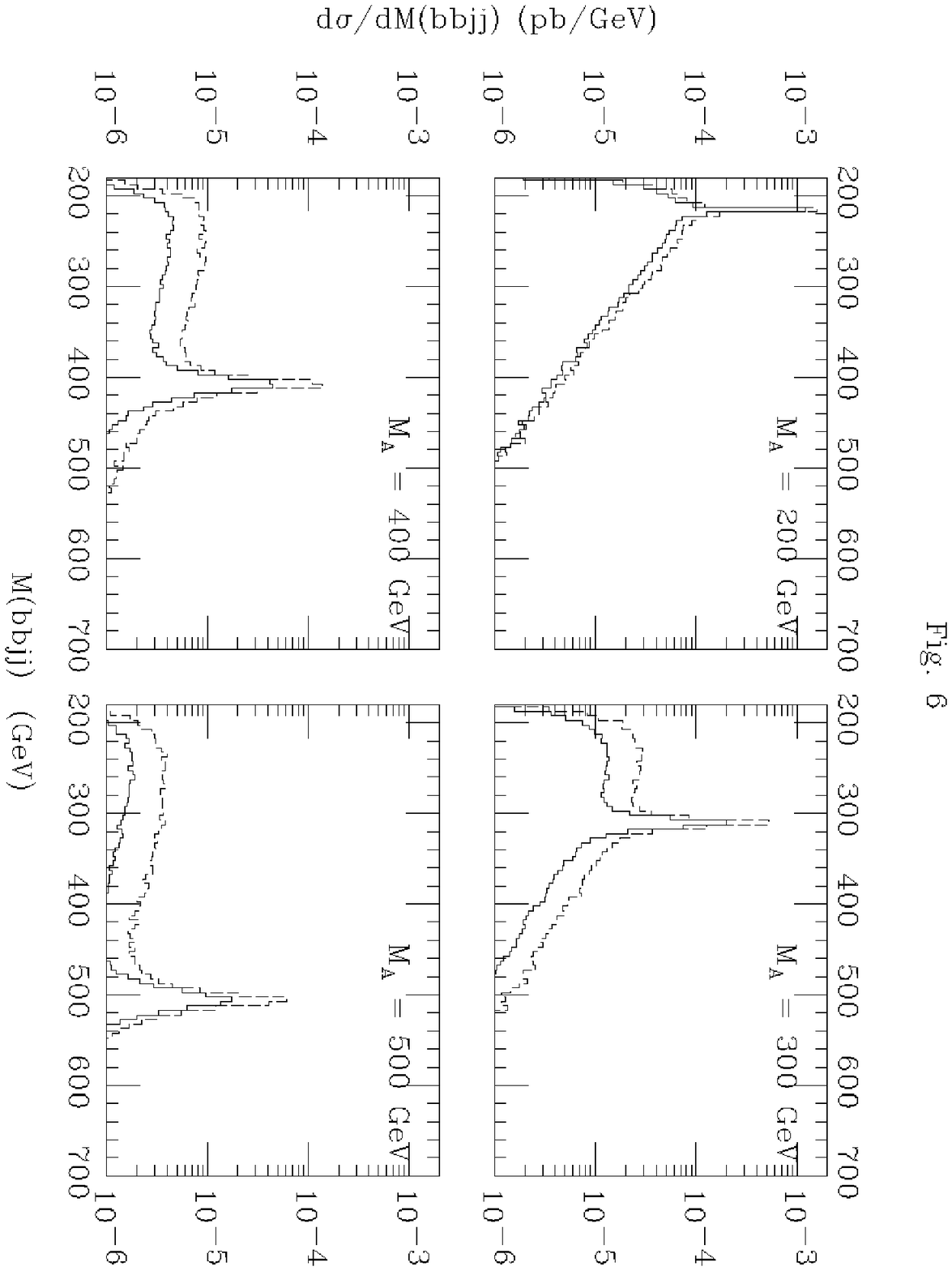,height=22cm}  
\vspace*{2cm}
\end{figure}
\clearpage
\begin{figure}[p]
~\epsfig{file=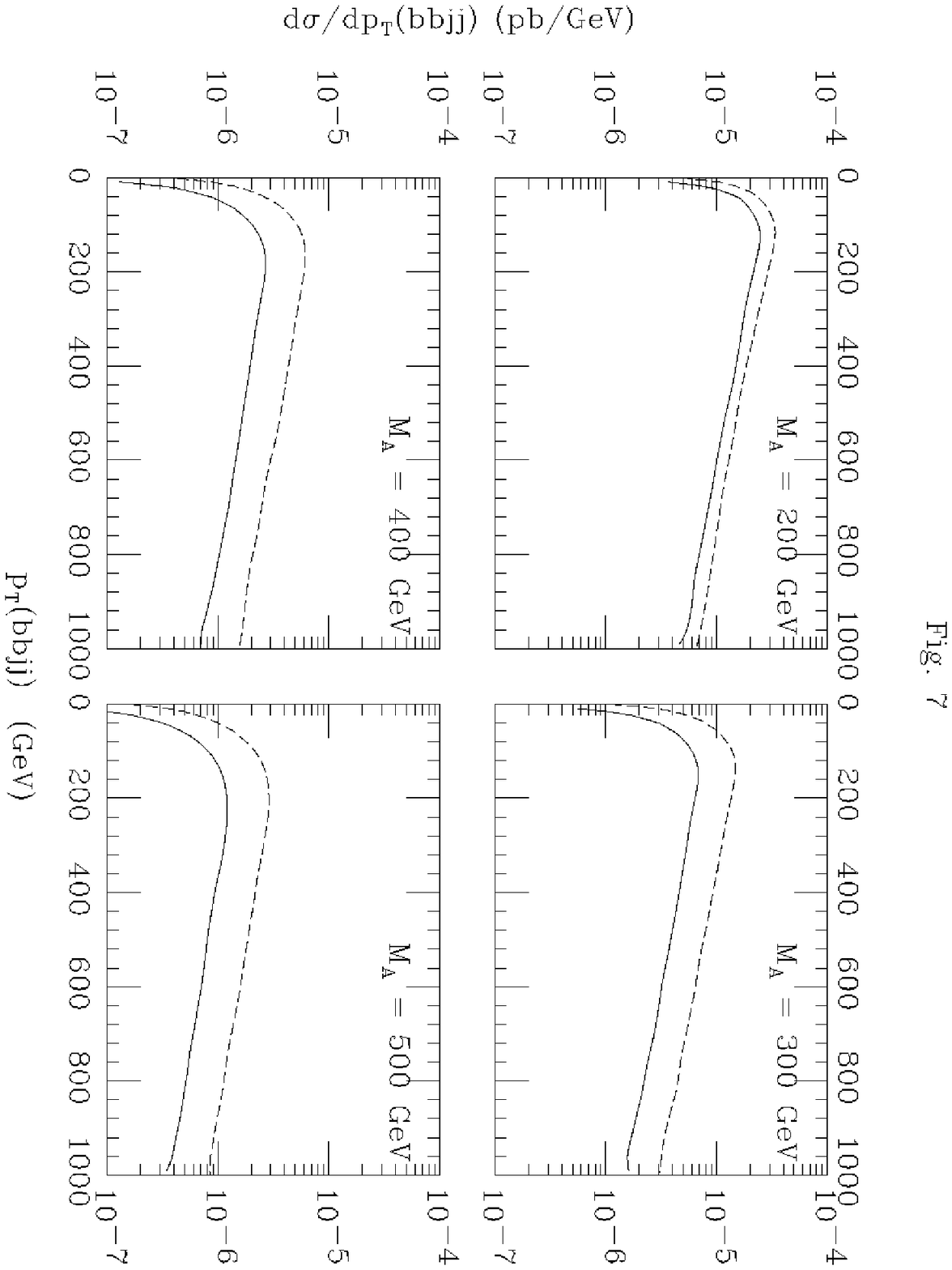,height=22cm}  
\vspace*{2cm}
\end{figure}
\vfill

\begin{thebibliography}{99}

\bibitem{CMS} CMS Technical Proposal, CERN/LHC/94-43 LHCC/P1 (December 1994).

\bibitem{ATLAS} ATLAS Technical Proposal,
CERN/LHC/94-43 LHCC/P2 (December 1994).

\bibitem{ioejames} S.~Moretti and W.J.~Stirling,
                    {\it Phys. Lett.} {\bf B347} (1995) 291; Erratum,
                    {\it ibidem}, {\bf B366} (1996) 451.

\bibitem{Pistarino} A. Ballestrero, E. Maina, S. Moretti and C. Pistarino,
                {\it Phys. Lett.} {\bf B320} (1994) 305.

\bibitem{bg} J.F.~Gunion, H.E. Haber, F. Paige, Wu-ki Tung and S.S.D. 
Willenbrock, \np B294 1987 621. 

\bibitem{KS} R.~Kleiss and W.J.~Stirling,
{\it Nucl. Phys.} {\bf B262} (1985) 235.

\bibitem{mana} C.~Mana and M.~Martinez,
{\it Nucl. Phys.} {\bf B287} (1987) 601.

\bibitem{ioPR} S. Moretti, \pr D50 1994 2016.

\bibitem{tim} T.~Stelzer and W.F.~Long, {\it Comp. Phys. Comm.} {\bf 81}
              (1994) 357.

\bibitem{HELAS} H.~Murayama, I.~Watanabe and K.~Hagiwara, HELAS: HELicity
                Amplitude Subroutines for Feynman Diagram Evaluations,
                {\it KEK Report} 91-11, January 1992.

\bibitem{FORM} J.A.M. Vermaseren, The Symbolic Manipulation 
Program FORM, Computer Algebra Netherland, Amsterdam 1991.

\bibitem{VEGAS} G.P.~Lepage, {\it Jour. Comp. Phys.} {\bf 27} (1978) 192.

\bibitem{corrMH0iMSSM} Y.~Okada, M.~Yamaguchi and
T.~Yanagida, {\it Prog. Teor. Phys. Lett.} {\bf 85} (1991) 1;\\
J.~Ellis, G.~Ridolfi and F.~Zwirner, {\it Phys. Lett.} {\bf B257} (1991) 83;
{\it Phys. Lett.} {\bf B262} (1991) 477;\\
H.E.~Haber and R.~Hempfling, {\it Phys. Rev. Lett.} {\bf 66} (1991) 1815;\\
R.~Barbieri and M.~Frigeni, {\it Phys. Lett.} {\bf B258} (1991) 395.

\bibitem{corrMHMSSM} A.~Brignole, J.~Ellis, G.~Ridolfi and F.~Zwirner,
                     {\it Phys. Lett.} {\bf B271} (1991) 123;\\
                     A.~Brignole, {\it Phys. Lett.} {\bf B277} (1992) 313.

\bibitem{0pmLEPLHCSSC} V.~Barger, K.~Cheung, R.J.~Phillips and A.L.~Stange,
                                  {\it Phys. Rev.} {\bf D46} (1992) 4914.

\bibitem{MARCIANO} W.J.~Marciano, \pr D29 1984 580.

\bibitem{widthtopMSSM} G.L.~Kane, Proceedings of the ``{\it Madison Workshop}''
                       (1979).

\bibitem{widthtopSM} J.H.~K\"uhn, {\it Act. Phys. Pol.} {\bf B12} (1981) 347;\\
                     J.H.~K\"uhn, {\it Act. Phys. Austr. Suppl. } {\bf XXIV}
                     (1982) 203.

\bibitem{topdecay} M. Je\.zabek and J.H.~K\"uhn,  \np B320 1989 20.

\bibitem{MRSA}  A.D.~Martin, R.G.~Roberts and W.J.~Stirling,
                \pr D50 1994 6734.

\bibitem{guide} See, for example:\\
J.F.~Gunion, H.E.~Haber, G.L.~Kane and S.~Dawson,
                {\it ``The Higgs Hunter Guide''}
                (Addison-Wesley, Reading MA, 1990)                 
and references therein.

\bibitem{them} A.~Djouadi, J.~Kalinowski and P.M.~Zerwas, \preprint\
                DESY 95--211, IFT--95--14, October 1995.

\end{thebibliography}
\end{document}